\documentclass[twocolumn,showpacs,preprintnumbers,amsmath,amssymb,showkeys]{revtex4-1}

\usepackage{graphicx}
\usepackage{dcolumn}
\usepackage{bm}
\usepackage{bezier}

\def\XXint#1#2#3{{\setbox0=\hbox{$#1{#2#3}{\int}$}
\vcenter{\hbox{$#2#3$}}\kern-.5\wd0}}

\begin{document}

\title{Persistent current noise in normal and superconducting nanorings}
\author{Andrew G. Semenov$^{1}$
and Andrei D. Zaikin$^{2,1}$
}
\affiliation{$^1$I.E.Tamm Department of Theoretical Physics, P.N.Lebedev
Physics Institute, 119991 Moscow, Russia\\
$^2$Institut f\"ur Nanotechnologie, Karlsruher Institut f\"ur Technologie (KIT),
76021 Karlsruhe, Germany
}

\begin{abstract}
We investigate fluctuations of persistent current (PC) in nanorings both
with and without dissipation and decoherence.
We demonstrate that such PC fluctuations may persist down to zero temperature
provided there exists either interaction with an external environment or an
external (periodic) potential produced, e.g., by quantum phase slips in superconducting nanorings.
Provided quantum coherence is maintained in the system PC noise remains coherent and
can be tuned by an external magnetic flux $\Phi_x$ piercing the ring.
If quantum coherence gets suppressed by interactions with a dissipative bath
PC noise becomes incoherent and $\Phi_x$-independent.

\end{abstract}

\pacs{to add}

\maketitle

\section{Introduction}

Persistent currents (PC) in normal and superconducting meso- and nanorings
is a fundamental ground state property of such systems. The basic physical reason
behind the existence of PC is quantum coherence of electron wave functions
which may survive at long distances provided the temperature remains low.
A lot is known about the average value of PC in such systems. E.g., in normal rings
this quantity was analyzed in details both theoretically \cite{thy} and experimentally \cite{exp}.
At the same time, only few studies of equilibrium fluctuations of PC are available.

While non-vanishing thermal fluctuations of PC can easily be expected, the issue
of quantum fluctuations of PC is somewhat less trivial since at
$T \to 0$ the system approaches its non-degenerate ground state.
Recently it was demonstrated \cite{SZ10} that in this case PC does not fluctuate
provided the current operator $\hat I$ commutes with the total Hamiltonian $\hat H$ of the
system. In all other cases PC fluctuates even though the system remains in its ground state. For
instance, quantum fluctuations of PC in mesoscopic rings may persist down to $T=0$ provided such
systems are coupled to an external dissipative bath \cite{Buttiker}.

Interestingly enough, quantum fluctuations of PC may also occur in the absence of
any dissipation. For instance, it is easy to verify that
the operators $\hat I$ and $\hat H$ do not commute for a quantum particle on a ring
with some (periodic) potential \cite{SZ10} and, hence, PC fluctuations do
not vanish even in the ground state at $T=0$. Since quantum coherence remains fully
preserved in this case the magnitude of PC fluctuations should depend on
the external magnetic flux \cite{SZ10}. It follows immediately that
by measuring the equilibrium current noise in meso- and nanorings it is possible
to effectively probe quantum coherence and decoherence in such systems.

The goal of this paper is to theoretically analyze persistent
current noise in both dissipative and non-dissipative systems.
In the presence of dissipation the time reversal symmetry is violated and PC noise
in metallic nanorings may be affected by decoherence. If, however, no source of dissipation is available,
the time reversal symmetry is preserved and PC fluctuations remain fully coherent, as we already
explained above. The first sutuation will be described within a model of a quantum particle
on a ring interacting with some quantum dissipative environment \cite{GZ98,Paco,GHZ} which
could be, e.g., a bath of Caldeira-Leggett oscillators or electrons in a disordered
conductor. Within this model PC fluctuations will be analyzed in section 2 both in the
perturbative and non-perturbative in the interaction regimes. The second situation will be
represented by a model of superconducting nanorings with quantum phase slips \cite{AGZ} which tend
to suppress PC in sufficiently large rings. PC noise in superconducting nanorings will be studied in
section 3. In section 4 we will briefly summarize our main observations.

\section{Particle on a ring in a dissipative environment}

\subsection{The model and effective action}

Let us consider a quantum particle with mass $M$ and electric charge $e$
moving in a 1d ring of radius $R$ pierced by magnetic flux $\Phi_x$. This quantum particle
interacts with some collective variable $V$ describing voltage fluctuations
in our dissipative bath. The total Hamiltonian for this system reads
\begin{equation}
   \hat H=\frac{(\hat \phi -\phi_x)^2}{2MR^2}+\hat H_{\rm env}(V)+\hat H_{\rm int}(\theta,V),
\label{H}
\end{equation}
where $\theta$ is the angle variable which controls the position of the particle
on the ring, $\hat \phi =-i\frac{\partial}{\partial\theta}$ defines the
angular momentum operator, $\phi_x=\Phi_x/\Phi_0$ and $\Phi_0=2\pi c/e$ is the
flux quantum (here and below we set the Planck's constant equal to unity $\hbar =1$).
The first term in Eq. (\ref{H}) is just the particle kinetic energy, $\hat H_{\rm env}(V)$
is the Hamiltonian of the bath, and
the term
\begin{equation}
\hat H_{\rm int}=e\hat V, \label{eV}
\end{equation}
accounts for Coulomb interaction between the particle and
the bath.

In what follows we will model a dissipative bath by a 3d
diffusive electron gas \cite{Paco,GHZ} with the inverse dielectric function
\begin{equation}
\frac{1}{\epsilon (\omega , k)}\approx \frac{-i\omega +Dk^2}{4\pi \sigma}.
\label{diel}
\end{equation}
Fluctuations of the electric potential $V$ in this dissipative environment are described by the correlator
\begin{equation}
\langle VV\rangle_{\omega , k}= -\coth \frac{\omega}{2T}{\rm Im}\frac{4\pi}{k^2\epsilon (\omega , k)}.
\end{equation}
Here $\sigma$ is the Drude conductivity of this gas, $D=v_Fl/3$
is the electron diffusion coefficient, $v_F$ is Fermi velocity
and $l$ is the electron elastic mean free path which is assumed to obey
the condition $k_F l \gg 1$ but to remain much
smaller than the ring radius $l \ll R$. We also point out that Eq.
(\ref{diel}) applies at not too high frequencies $\omega\ll \omega_c \sim v_F/l$.

Employing the definition for the current operator
\begin{equation}
    \hat I=\frac{e}{2\pi}\dot{\hat \theta}=\frac{ie}{2\pi}[\hat H,\hat\theta ]=\frac{e(\hat\phi -\phi_x)}{2\pi MR^2}
\label{curop}
\end{equation}
and making use of the Heisenberg representation
$\hat I(t)=e^{it\hat H}\hat I e^{-it\hat H}$,
we introduce the current-current correlation function
$\langle\hat I(t)\hat I(0)\rangle$ and define PC noise power \cite{SZ10}
\begin{equation}
S(t)=\frac12\langle\hat I(t)\hat I(0)+\hat I(0)\hat I(t)
\rangle-\langle\hat I\rangle^2=\int\frac{d\omega}{2\pi}S_\omega e^{-i\omega t}.
\label{sdef}
\end{equation}
In order to evaluate this correlation function we further
introduce the evolution operator
$\hat U(t,t_0)$ and define the density matrix operator
$\hat \rho(t)=\hat U(t,0)\hat\rho_i \hat U^\dag(t,0)$,
where $\rho_i$ is the initial density matrix. Since our goal here is to analyze quantum dynamics of the particle rather than that of
the bath, it will be convenient to employ the standard influence functional technique \cite{FH} and trace out fluctuating potential $V$.
Making use of a simplifying assumption that at the initial time moment the total density matrix is factorized into the product
of the equilibrium bath density matrix and that of a particle $\hat \rho_i$,
one can rewrite the evolution equation for the density matrix in the form of a double path integral
over the angle variables $\theta^F$ and $\theta^B$ defined respectively on
the forward and backward parts of the Keldysh contour
\begin{widetext}
\begin{eqnarray}
  \rho(\theta_1,\theta_2;t)=\sum\limits_{m_1,m_2=-\infty}^\infty e^{i(\theta_1+2\pi m_1)\phi_x-i(\theta_2+2\pi m_2)\phi_x} \int\limits_0^{2\pi}d\theta_1'd\theta_2' e^{-i(\theta_1'-\theta_2')\phi_x}\rho_i(\theta_1',\theta_2')\nonumber\\
  \times\int\limits_{\theta^F(0)=\theta_1'}^{\theta^F(t)=\theta_1+2\pi m_1}\mathcal D \theta^F\int\limits_{\theta^B(0)=\theta_2'}^{\theta^B(t)=\theta_2+2\pi m_2}\mathcal D \theta^B e^{i\int\limits_{0}^{t}[((\dot \theta^F)^2-(\dot \theta^B)^2)/4E_C]dt'}e^{-iS_R-S_I},
\end{eqnarray}
where $\rho(\theta_1,\theta_2;t)\equiv\langle\theta_1|\hat\rho(t)|\theta_2\rangle$, $E_C=1/(2MR^2)$  and $\exp(-iS_R-S_I)$
is the influence functional.
Calculation of this functional amounts to averaging over the quantum variable $V$ which is also defined on the Keldysh contour.
Such averaging was performed, e.g., in Refs. \onlinecite{GZ1,GZS,GZ2} for a degenerate electron gas where Pauli exclusion
principle should explicitly be accounted for.
The same procedure can be employed in our present situation of a particle on a ring where no Pauli principle needs
to be included.  Introducing the new variables  $\theta_+=(\theta^F+\theta^B)/2$ and $\theta_-=\theta^F-\theta^B$,
after the standard algebra we obtain
\begin{equation}
S_{R}[\theta_{+},\theta_-]=\pi\alpha \sum\limits_{n=1}^\infty a_n n \int\limits_{0}^{t} dt' \dot \theta_{+}(t')\sin(n\theta_-(t')),
\label{inffunc1b}
\end{equation}
and
\begin{equation}
  S_{I}[\theta_{+},\theta_-]=-2\pi\alpha \sum\limits_{n=1}^\infty a_n\int\limits_{0}^{t} dt' \int\limits_{0}^{t} dt''\frac{\pi T^2}{\sinh^2(\pi T(t'-t''))}\cos(n(\theta_{+}(t')-\theta_{+}(t'')))\sin\frac{n\theta_-(t')}{2}\sin\frac{n\theta_-(t'')}{2},
  \label{inffunc1a}
\end{equation}
\end{widetext}
where $\alpha=3/(8 k_F^2l^2)$ is the effective coupling constant in our problem and $a_n$ are the Fourier coefficients equal
to $a_n=(2/(\pi r))\ln(r/n)$ for $n<r\equiv R/l \gg 1$ and to zero $a_n=0$ otherwise. The weak disorder condition $k_Fl \gg 1$ implies
a small effective coupling constant $\alpha \ll 1$. It is also worth pointing out that the above influence functional reduces
to one derived within the Caldeira-Leggett model provided one chooses $\alpha =\eta R^2/\pi$ and $a_n=\delta_{1n}$, where
$\eta$ defines effective friction produced by the bath and $\delta_{ij}$ is the Kronecker symbol.

In the case of the Caldeira-Leggett environment decoherence of
a quantum particle on a ring was investigated with the aid of
both real-time \cite{GZ98} and Matsubara \cite{Paco} techniques which
yield similar results, i.e. exponential suppression of quantum
coherence down to $T=0$ at sufficiently large ring radii.
This result is by no means surprizing since the model is
exactly equivalent to that of Coulomb blockade in a
single electron box where exponential reduction of the effective charging energy at
large conductances is also well established \cite{pa91,HSZ}.
The model of a particle in a diffusive electron gas
was employed by different authors
\cite{Paco,GHZ,HlD,CH,KH,SZ09} investigating
the effect of interaction-induced decoherence on the average
value of PC. Below we will make use of this model in order
to analyze PC fluctuations in the presence of quantum
decoherence.

\subsection{Perturbation theory}
Provided the ring radius $r$ is sufficiently small one can proceed perturbatively in $\alpha$. Consider the kernel
of the evolution operator $\mathcal U$ which establishes a relation between the density matrix elements at different moments of time
\begin{equation}
\tilde \rho (m_1,m_2;t)=\sum\limits_{m_1',m_2'}\mathcal U_{m_1,m_1'}^{m_2,m_2'}(t)\tilde \rho (m_1',m_2';0),
\end{equation}
where
\begin{equation}
\tilde \rho (m,n;t)=\int\limits_0^{2\pi}\frac{d\theta_1}{2\pi}\int\limits_0^{2\pi}\frac{d\theta_2}{2\pi}\rho (\theta_1,\theta_2;t) e^{-im\theta_1+in\theta_2}
\end{equation}
is the density matrix in the momentum representation which remains diagonal for the problem in question.
Therefore one can rewrite the above evolution equation as a matrix one for the diagonal elements of the density matrix.

\begin{figure}[t]
\includegraphics[width=0.9\columnwidth]{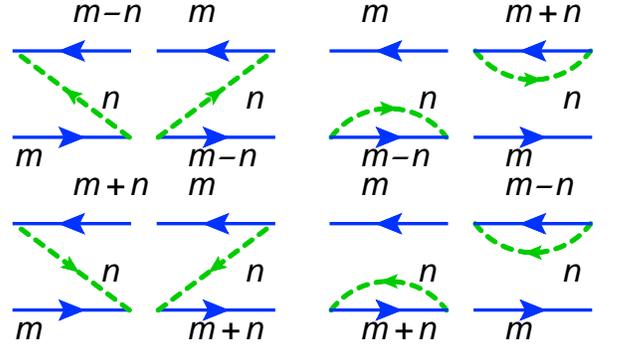}
\caption{First order self-energy diagrams}
\label{fig1}
\end{figure}

In order to account for interaction effects we will employ the perturbation theory similar to that developed for the Coulomb blockade problem
\cite{GZ94,SS}. Explanding the influence functional $\exp(-iS_R-S_I)$ in powers of the coupling constant $\alpha$
and taking into account first order diagrams depicted in Fig. \ref{fig1} we arrive at the evolution kernel
\begin{equation}
\tilde{\mathcal U}_{m,m'}^{m,m'}(\omega)\equiv\int\limits_0^\infty dt e^{i\omega t}\mathcal U_{m,m'}^{m,m'}(t)=\left[\frac{i}{\omega+i\tilde\Gamma_\omega}\right]_{m,m'},
\end{equation}
where
\begin{multline}
[\tilde \Gamma_\omega]_{m+n,m}=-\frac{\pi\alpha a_{|n|}}{2}\left(\frac{E_{m+n}-E_m+\omega}{e^{\frac{E_{m+n}-E_m+\omega}{T}}-1}\right.\\
\left.+\frac{E_{m+n}-E_m-\omega}{e^{\frac{E_{m+n}-E_m-\omega}{T}}-1}\right),
\end{multline}
\begin{equation}
[\tilde \Gamma_\omega]_{m,m}=-\sum\limits_{n=1}^\infty\left([\hat \Gamma_\omega^{(0)}]_{m+n,m}+ [\hat \Gamma_\omega^{(0)}]_{m-n,m}\right).
\end{equation}

Within the same approximation for PC noise power one finds
\begin{equation}
S_\omega=2\Re\sum\limits_{m,n}I_m\left(\tilde{\mathcal U}_{m,n}^{m,n}(\omega)-\frac{iP_m}{\omega+i0}\right)I_nP_n
\end{equation}
where $I_n=2E_C(m-\phi_x)$ and $P_n=e^{-E_C(n-\phi_x)^2/T}/\mathcal Z$ define respectively the current and the distribution function in the absence of interactions. The latter quantity also involves the partition function for our system  $\mathcal Z=\sum_n e^{-E_C(n-\phi_x)^2/T}$.

\begin{figure}[t]
\includegraphics[width=0.9\columnwidth]{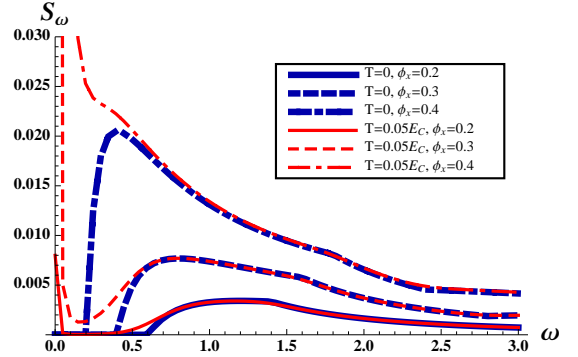}
\caption{PC noise power at  $\pi\alpha =0.05$, $r=5$ at different values of $T$ and $\phi_x$. The frequency $\omega$ and noise power $S_\omega$ are normalized respectively by $E_C$ and by $e^2E_C/(4\pi^2)$.}
\label{fig2}
\end{figure}

We have numerically evaluated both the kernel of the evolution operator and PC noise power. Our results are
displayed in Fig. \ref{fig2} at different values of temperature and the magnetic flux.
One observes a strong dependence of PC noise power on the external magnetic flux $\phi_x$. This feature
indicates the coherent nature of PC noise \cite{SZ10} which clearly persists also in the presence of dissipation provided the effect of the latter is
sufficiently weak. It turns out that PC noise grows with increasing $\phi_x$ and formally diverges in the vicinity of the point $\phi_x=0.5$.
This divergence has to do with the fact that the distance between the two lowest energy levels $\delta E(\phi_x)=E_C(1-2|\phi_x|)$
becomes small in this limit. Hence, the system undergoes rapid transitions between these states corresponding to different PC values.
As a result, PC fluctuations in our system get greatly enhanced as soon as the flux approaches the value $\phi_x=0.5$.

We also emphasize that PC noise does not vanish even in the limit $T \to 0$. In this case $S_\omega$ equals to zero only at frequencies
below the inter-level distance $\omega <\delta E(\phi_x)$ and remains non-zero at higher values of $\omega$. We also point out that PC noise
vanishes at $\phi_x=0$ if evaluated perturbatively in the lowest order in $\alpha$. However, non-zero PC noise at $\phi_x=0$ is recovered
if one goes beyond perturbation theory in the interaction, as it will be demonstrated below. The latter property also follows from
the general expressions formulated in terms of the exact eigenstates of the total Hamiltonian \cite{SZ10}.

Finally, we observe that at non-zero $T$ there appears additional zero frequency peak in $S_\omega$. This peak grows with increasing temperature and eventually merges with all other peaks forming a wide hump at
sufficiently high $T$.  In this case quantum coherence gets essentially suppressed and PC noise becomes flux-independent.

\subsection{Non-perturbative regime}

Let us now go beyond perturbation theory in the effective coupling constant $\alpha$.
At the first sight this step might be considered unnecessary since within the applicability range of our model
this coupling constant always remains small $\alpha \ll 1$. However, it turns out \cite{GHZ} that the actual parameter that controls the strength
of interaction effects is $\alpha r$ rather than $\alpha$. Hence, should the ring radius be sufficiently large, i.e.
\begin{equation}
4\pi\alpha r \gg 1,
\label{npl}
\end{equation}
perturbation theory in the interaction fails and non-perturbative analysis of the problem becomes inevitable.

\begin{figure}[t]
\includegraphics[width=0.97\columnwidth]{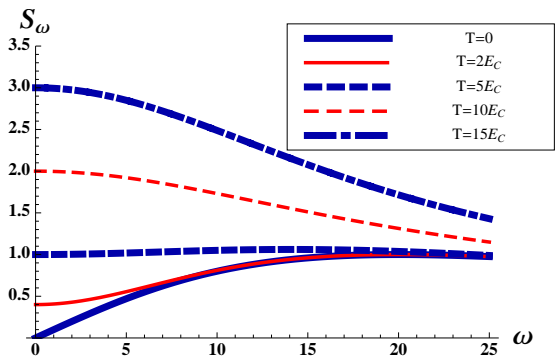}
\caption{Noise power at different temperatures for $\pi\alpha =0.05$ and $r=10$. Frequency $\omega$ and PC noise power are normalized respectively by $E_C$ and by $e^2E_C/(4\pi^2)$.}
\label{f5}
\end{figure}

In the limit (\ref{npl}) and not too low temperature one can employ the semiclassical approximation which amounts to expanding the action
(\ref{inffunc1b}), (\ref{inffunc1a}) up to quadratic in $\theta_{-}$ terms. The resulting effective action can be exactly
reformulated in terms of the quasiclassical Langevin equation \cite{Schmid,AES,GZ92}
for the "center-of-mass" variable $\theta_+$. For the model under consideration this equation reads
\begin{eqnarray}
-\frac{1}{2E_C}\ddot \theta_+(t)-\frac{\gamma}{2}\dot \theta_+(t)=\sum\limits_{n=1}^\infty (\xi_n(t)\cos(n\theta_+(t))\quad\nonumber\\+\lambda_n(t)\sin(n\theta_+(t))),
\label{langev}
\end{eqnarray}
where we introduced the parameter
\begin{equation}
\gamma=2\pi\alpha \sum\limits_{n=1}^\infty a_n n^2=4\pi\alpha r^2
\end{equation}
and defined Gaussian stochastic fields $\xi_n(t)$ with the correlators
\begin{eqnarray}
\langle\xi_n(t)\xi_m(t')\rangle_{\xi,\lambda}=\langle\lambda_n(t)\lambda_m(t')\rangle_{\xi,\lambda}=\qquad\nonumber\\=-\delta_{m,n}\pi\alpha a_n n^2\frac{\pi T^2}{\sinh^2(\pi T(t-t'))},
\label{cor1}
\end{eqnarray}
\begin{equation}
\langle\xi_n(t)\lambda_m(t')\rangle_{\xi,\lambda}=0.
\label{cor2}
\end{equation}
At high temperatures the white noise limit is realized,
\begin{equation}
\langle\xi_n(t)\xi_m(t')\rangle_{\xi,\lambda}=2\delta_{m,n}\pi\alpha a_n n^2T\delta(t-t'),
\label{wn}
\end{equation}
and Eq. (\ref{langev}) can be solved exactly.  In this limit we obtain
\begin{equation}
  S_\omega=\frac{e^2\gamma T E_C^2}{\pi^2(\omega^2+(\gamma E_C)^2)}.
\label{htpn}
\end{equation}
At lower values of $T$ the approximation (\ref{wn}) fails and more accurate Eqs. (\ref{cor1}), (\ref{cor2}) should be employed.
Treating the noise terms in Eq. (\ref{langev}) perturbatively \cite{GZ92} and taking into account only the zeroth and the first order
contributions we arrive at the result
\begin{equation}
\theta_+(t)=\theta_+^{(0)}+\theta_+^{(1)}(t),
\label{fior}
\end{equation}
where $\theta_+^{(0)}$ is some physically irrelevant constant and $\theta_+^{(1)}(t)$ obeys the equation
\begin{eqnarray}
-\frac{1}{2E_C}\ddot \theta_+^{(1)}(t)-\frac{\gamma}{2}\dot \theta_+^{(1)}(t)=\sum\limits_{n=1}^\infty \xi_n(t),
\label{langev1}
\end{eqnarray}
which allows to immediately recover the noise power
\begin{equation}
S_\omega=\frac{e^2\gamma E_C^2}{2\pi^2(\omega^2+(\gamma E_C)^2)}\omega\coth\frac{\omega}{2T},
\label{noiseHT}
\end{equation}
Obviously, this result reduces back to Eq. (\ref{htpn}) in the limit $T \gg \omega$.
At $\omega \ll \gamma E_C$ the parameter $E_C$ drops out from the expression for the noise power and we get
\begin{equation}
    S_\omega =\frac{e^2\omega}{2\pi^2\gamma}\coth\frac{\omega}{2T},
  \end{equation}
i.e. in this case $S_\omega \propto 1/\alpha$. For $ \omega \to 0$ this expression further reduces to $S_0 \propto T/\gamma$.
The function $S_\omega$ (\ref{noiseHT}) is displayed in Fig. \ref{f5} at different temperatures.

Comparing these results with those derived perturbatively in Sec. 2.B we observe that, while at weak interactions
the PC noise remains coherent and, hence, depends on the external magnetic flux $\Phi_x$, in the limit of strong interactions (\ref{npl})
this dependence is practically absent. This is due to strong decoherence effect produced by our dissipative bath. As a result, in the limit
(\ref{npl}) the average value of PC gets exponentially suppressed, while PC noise does not vanish but becomes essentially incoherent.

Note that, strictly speaking, the flux-dependent contribution to $S_\omega$ survives also in this limit, but it remains exponentially small, as
it is demonstrated by the analysis \cite{SZ11}. Technically, the presence of such $\Phi_x$-dependent correction to the result (\ref{noiseHT})
is related the fact that the angle variable $\theta$ is defined on a ring, i.e. is compact. The Langevin equation approach employed here
"decompactifies" this variable, thereby capturing only the $\phi_x$-independent contributions to PC noise power.
In order to estimate the leading $\phi_x$-dependent correction to Eq. (\ref{noiseHT}) one can follow the analysis initially developed
for the problem of weak Coulomb blockade in metallic quantum dots \cite{GZ96}.
This approach allows to obtain the relation between the density matrices and expectation values evaluated for the problems described by
the same Hamiltonian but respectively compact and non-compact variables. Without going into corresponding details here we only quote the
result \cite{SZ11}
\begin{eqnarray}
S_\omega=\frac{e^2\gamma E_C^2\omega\coth(\omega/2T)}{2\pi^2(\omega^2+(\gamma E_C)^2)}\qquad\qquad\qquad\qquad\nonumber\\
\times\left(1-A e^{-4\pi \alpha r}\cos(2\pi\phi_x)\right),
\label{noiseTot}
\end{eqnarray}
where $A \propto \alpha r$. Thus, in the non-perturbative limit (\ref{npl}) the coherent flux-dependent contribution is indeed
exponentially small and can be safely neglected as compared to the main incoherent term (\ref{noiseHT}).

\section{PC noise in thin superconducting nanorings}

Let us now turn to the analysis of persistent current noise in a dissipativeless system which will be a superconducting nanoring.
As long as the ring remains sufficiently thick superconducting fluctuations can be ignored and, hence, there exists no physical mechanism
that could cause PC fluctuations. If, however, the ring becomes thin, superconductivity may be disrupted in various places in the ring
due to fluctuations of the order parameter. At low temperatures most important fluctuations of that kind are quantum phase slips (QPS) \cite{AGZ}.
Below we will demonstrate  that the properties of superconducting nanorings in the presence of QPS are described by the effective theory
equivalent to that for a quantum particle on a ring in a periodic potential.

The starting point of our derivation is the expression for the grand partition function $\mathcal  Z$.
This expression can be represented in terms of a path integral over the phase $\varphi$ of the superconducting order parameter.
Employing the low-energy effective action for a quasi-1d superconducting wire \cite{ZGOZ,GZ01,anne} one finds
\begin{equation}
\mathcal  Z=\sum\limits_{m,n}\int \mathcal D\varphi e^{-\frac{\lambda}{2\pi}\int dxd\tau\left(v(\partial_x\varphi)^2+v^{-1}(\partial_\tau\varphi)^2\right)},
\label{partf}
\end{equation}
where we defined $\lambda =\frac{\pi^{2} N_{0}D\Delta s}{2v}$, $s$ is the ring cross section, $N_{0}$ is the density of states at Fermi level, $\Delta$ is
the absolute value of the superconducting order parameter, $D$ is the diffusion coefficient and $v=\sqrt{\pi\sigma \Delta s/C}$ is the velocity
of low energy plasmon mode propagating along the wire (the so-called Mooji-Sch\"on mode). Here $\sigma$ is the Drude normal state conductivity of our metallic ring and
$C$ is the wire capacity per unit length.

The path integral in Eq. (\ref{partf}) should be performed with periodic (in imaginary time) boundary conditions
$\varphi(x,0)=\varphi(x,\beta)+2\pi m$, where $m$ is an arbitrary integer number (the so-called winding number).
The boundary conditions should also be periodic with respect to the spatial coordinate along the ring and, in addition, should
be sensitive to the magnetic flux piercing the ring, i.e. $\varphi(L,\tau)=\varphi(0,\tau)+2\pi(\phi_x+n)$.
Here $L=2\pi R$ is the ring perimeter, $\phi_{x}=\Phi/\Phi_{sc0}$ and $\Phi_{sc0}=\Phi_0/2$ is the superconducting flux quantum.

The partition function (\ref{partf}) can be evaluated semiclassically. As usually, in the main approximation it suffices
to take into account all relevant saddle point configurations of the phase variable $\varphi$ which satisfy the equation
\begin{equation}
(\partial_\tau^2+v^2\partial_x^2)\varphi(x,\tau)=0.
\label{speq}
\end{equation}
Apart from trivial solutions of this equation linear in $\tau$ and $x$
there exist nontrivial ones which correspond to virtual phase jumps by $\pm 2\pi$ at various points of a superconducting ring.
These quantum topological objects can be viewed as vortices in space-time and represent QPS events \cite{AGZ,ZGOZ,GZ01}.
All these configurations can be effectively summed up with the aid of the approach involving the so-called duality transformation.

In order to proceed let us express the general solution of Eq. (\ref{speq}) in the form
\begin{equation}
  \varphi(x,\tau)=a_{m}\tau+b_{n}x+\varphi^{qps}(x,\tau),
\end{equation}
where $a_m$ and $b_n$ are some constants fixed by the boundary conditions. We also introduce the vorticity field $\varpi(x,\tau)$ with the aid of the relations
\begin{equation}
 v\partial_x\varpi=\partial_\tau\varphi^{qps}\quad v\partial_\tau\varpi=-v\partial_x\varphi^{qps}.
\end{equation}
This field is single-valued and it obeys the equation
\begin{equation}
\partial^2_\tau\varpi+v^2\partial^2_x\varpi=2\pi v\sum\limits_j\nu_j\delta(x-x_j)\delta(\tau-\tau_j),
\end{equation}
where $x_j$, $\tau_j$ and $\nu_j$ denote respectively the space and time coordinates  of the $j$-th phase slip and its topological charge (phase winding). The partition function for a given saddle point solution can be rewritten as a path integral over the $\varpi$-field containing the functional delta function which follows from the above equation. Performing a summation over all possible QPS configurations we obtain
\begin{widetext}
\begin{multline}
\mathcal Z=\sum\limits_{N=0}^\infty\sum\limits_{\nu_1,..,\nu_N=\pm1}\int\limits_0^{2\pi}\frac{dz}{2\pi}\int dx_1d\tau_1...dx_Nd\tau_N\sum\limits_{m,n=-\infty}^\infty e^{2\pi in\phi_x-\frac{\pi v\beta m^2}{2g L}-\frac{\pi L n^2}{2g \beta v}} \left(\frac{\gamma_{QPS}}{2}\right)^N e^{2\pi i m\sum\limits_j\nu_j\frac{x_j}{L} -2\pi i n\sum\limits_j\nu_j\frac{\tau_j}{\beta}}
\\\int\mathcal D\varpi
e^{-\frac{g}{2\pi}\int dxd\tau\left(v(\partial_x\varpi)^2+v^{-1}(\partial_\tau\varpi)^2\right)}\ e^{iz\sum\limits_j\nu_j}
\delta\left(\partial^2_\tau\varpi+v^2\partial^2_x\varpi-2\pi v\sum\limits_j\nu_j\delta(x-x_j)\delta(\tau-\tau_j)\right).
\label{zgen}
\end{multline}
\end{widetext}
Here $\beta =1/T$
\begin{equation}
 \gamma_{QPS} \sim (g_\xi \Delta /\xi )\exp (-a g_\xi )
\label{gqps}
\end{equation}
is the QPS rate \cite{GZ01},  $g_\xi = 4\pi N_0Ds/\xi $ is the dimensionless conductance of the wire segment of length equal to the
superconducting coherence length $\xi$ and $a$ is a numerical prefactor of order one. Eqs. (\ref{zgen}), (\ref{gqps}) are applicable
provided $g_\xi$ is sufficiently large, i.e. $g_\xi e^{-ag_\xi /2}\ll 1$.

Rewriting the delta function in Eq. \ref{zgen}) as a path integral of the exponent and performing summation over all
QPS configurations as well as integration over $\varpi$ we get
\begin{equation}
 \mathcal Z =\sum\limits_{m,n=-\infty}^\infty e^{2\pi in\phi_x}\int\mathcal D\theta e^{-S_{\rm eff}[\theta ]},
\end{equation}
where
\begin{multline}
S_{\rm eff}=\int dxd\tau\left(\frac{(\partial_\tau\theta )^2+v^2(\partial_x\theta )^2}{8\pi v \lambda}-\gamma_{QPS}\cos\theta \right).
\label{efacchi}
\end{multline}
In contrast to the original problem here the path integration is performed over the single-valued field $\theta $ with periodic boundary conditions
 \begin{equation}
  \theta (x,\beta)-\theta (x,0)=2\pi n\quad \theta (L,\tau)-\theta (0,\tau)=2\pi m.
\end{equation}

PC noise power can be evaluated directly making use of the above equations and the expression for the current $I$, which is
just proportional to the phase difference around the ring,
\begin{equation}
I(\tau)=\frac{2\pi e v\lambda}{L}\left(\varphi(L,\tau)-\varphi(0,\tau)\right).
\end{equation}
Having evaluated the Matsubara current-current correlation function one performs analytic continuation to real times and,
taking into account the fluctuation-dissipation theorem, arrives at PC noise power spectrum $S_\omega$.

In the limit of low temperature $T \ll v/L$  and provided the ring perimeter is not too large
one can ignore the spatial dependence of the field $\theta$ and, hence neglect the term $v^2(\partial_x\theta)^2$ in the effective action (\ref{efacchi}).
After that our problem becomes effectively zero-dimensional and exactly equivalent to that of a particle on a ring in the presence of the cosine
external potential. In other words, we have mapped our problem onto that described by the Hamiltonian
\begin{equation}
   \hat H=\frac{(\hat \phi -\phi_x)^2}{2MR^2}+U_0 (1-\cos (\kappa \theta )) ,
\label{H1}
\end{equation}
where one should now identify $\kappa=1$,
\begin{equation}
\frac{1}{MR^2}\to E_R\equiv\frac{\pi^2 N_0 D\Delta s}{R}\sim \frac{g_{\xi}\Delta\xi}{R}
\end{equation}
and
\begin{equation}
U_0\to 2\pi R\gamma_{QPS} \sim \frac{g_{\xi}\Delta R}{\xi} e^{-ag_\xi}.
\end{equation}
The analysis of PC fluctuations for the model (\ref{H1}) in the limit of strong external potential $U_0 \gg E_R$ was carried
out in Ref. \onlinecite{SZ10}. In the case of superconducting nanorings it yields
\begin{multline}
 S_\omega=\frac{e^2\Omega^3}{2\pi U_0} \left(\delta(\omega-\Omega-\Lambda\cos(2\pi\phi_{x}))\right.\\\left.+\delta(\omega+\Omega+\Lambda\cos(2\pi\phi_{x}))\right),
\label{scnol}
\end{multline}
where $\Omega=\pi\sqrt{\pi N_0 D\Delta s \gamma_{qps}} \sim g_{\xi}\Delta e^{-ag_{\xi}/2}$ is the frequency of oscillations of the ''particle''  near the bottom of cosine potential and
\begin{equation}
\Lambda=256\sqrt{\frac{U_{0}^{3}}{\pi\Omega}}e^{-\frac{8U_{0}}{\Omega}}.
\end{equation}
The result (\ref{scnol}) demonstrates that in the low temperature limit PC noise power spectrum $S_\omega$ differs from zero due to the effect of
QPS and has the form of two sharp peaks
at frequencies well below the superconducting gap $\Delta$. The exact positions of these peaks can be tuned by the flux $\phi_x$ piercing the ring,
though only weakly, since in this case $\Omega \gg \Lambda$. Note that the condition $U_0 \gg E_R$ is equivalent to $R \gg R_c \sim \xi \exp (ag_\xi /2)$
in which case the average value of PC is exponentially suppressed \cite{AGZ} $\langle I\rangle \propto \exp (-R/R_c)$. The PC noise power spectrum peaks
also remain small in this case, though they decrease only as $S_\omega \propto R_c/R$ with increasing ring radius.

In the opposite case $E_R\gg U_0$ the cosine potential term remains small as compared to the particle kinetic energy. Hence,
in this limit the effect of QPS may be considered perturbatively in $\gamma_{QPS}$. In the absence of QPS the noise power $S_\omega$
shows only one peak at zero frequency with the amplitude equal to the current dispersion, i.e.
\begin{equation}
 S_\omega=2\pi (\langle \hat I^2\rangle-\langle\hat I\rangle^2)\delta(\omega).
\end{equation}
The amplitude of this peak decreases with temperature and tends to zero in the limit $T \to 0$. Employing the standard quantum mechanical perturbation theory,
in the lowest non-vanishing order in $U_0$ one recovers additional peaks at frequencies corresponding to the transitions between neighboring energy levels.
These peaks survive even at zero temperature $T=0$ in which case one finds
\begin{widetext}
\begin{multline}
S_{\omega}=\frac{e^{2}U_{0}^{2}}{\pi (1+2\phi_{x})^{2}}\left(\delta(\omega-E_{R}(1/2+\phi_{x}))+\delta(\omega+E_{R}(1/2+\phi_{x}))\right) \\ +\frac{e^{2}U_{0}^{2}}{\pi (1-2\phi_{x})^{2}}\left(\delta(\omega-E_{R}(1/2-\phi_{x}))+\delta(\omega+E_{R}(1/2-\phi_{x}))\right).
\end{multline}
\end{widetext}
This equation is applicable as long as the condition
\begin{equation}
\xi g_\xi \ll R\ll\min\left(R_c,\frac{v}{2\pi T}\right)
\end{equation}
remains fulfilled. We observe that in this case PC noise power $S_\omega$ has the form of four sharp peaks at frequencies which strongly depend on the external flux $\phi_x$.
E.g., by tuning the flux to the value close to one half of the superconducting flux quantum $\phi_x \approx \pm 1/2$ one should observe strong enhancement of
noise peaks which occur in the vicinity of zero frequency $\omega =0$. The physical reason for this enhancement is the same as that already
discussed in Sec. 2B: The energies of two lowest levels become close to each other at such values of $\phi_x$ implying the
possibility of rapid transitions between these states. Such intensive transitions, in turn, imply strong fluctuations of PC.
Exactly at resonance $\phi_x = \pm 1/2$ the second order perturbation theory fails and more accurate treatment becomes necessary.

\section{Conclusions}

In this paper we analyzed fluctuations of persistent current in nanorings with and without dissipation.
Specifically, we restricted our attention to PC noise and evaluated symmetric current-current correlation function.
Comparing the results obtained within two different models analyzed in Sec. 2 and 3 we observe both similarities and
important differences in the behavior of these systems.

To begin with, in the absence of interactions and dissipation in the model of a particle on a ring (Sec. 2) as well as in the absence of phase slips
in superconducting nanorings (Sec. 3) PC fluctuates only at non-zero $T$ and
no such fluctuations could occur provided the system remains in its ground state at $T=0$.
In the presence of interactions in the first model or quantum phase slips in the second model
the current operator does not anymore commute with the total Hamiltonian of the system and
fluctuations of PC do not vanish down to zero temperature. Yet another qualitative similarity
between these systems is that in both cases PC noise decreases with increasing the ring radius $R$.

The most important physical difference between the models considered in Sec. 2 and 3 is the presence of dissipation and, hence, decoherence
in the first model and their total absence in the second model. Accordingly, at low temperatures
PC noise always remains coherent in the second case which implies
that PC noise power spectrum essentially depends on the magnetic flux $\Phi_x$ piercing the ring. In the absence of dissipation at $T \to 0$ PC noise
has the form of sharp peakes at frequencies corresponding to energy differences between the system states for which quantum mechanical transitions
are possible. At flux values close to one half of the flux quantum some energy levels also become close to each other which means strong
enhancement of PC fluctuations.

Coherent fluctuations of PC are also possible in the presence of dissipation provided its effect remains sufficiently weak
and the ring radius remains small. In this limit decoherence effect of the external dissipative bath is still insignificant.
Narrow peaks in PC noise get somewhat broadened even at $T \to 0$ due to the presence of dissipation, but the dependence of $S_\omega$ on the magnetic flux
persists also in this case. In rings with larger radii, on the contrary, fluctuations in the dissipative bath strongly suppress quantum
coherence down to $T=0$ and induce incoherent $\Phi_x$-independent
current noise in the ring which persists even at $\Phi_x=0$ when the average PC is absent.
Thus, quantum coherence and its suppression by interactions in meso- and nanorings can be
experimentally investigated by measuring PC noise and its dependence on the external
magnetic flux. It would be interesting to carry out such experiments in the near future.


\begin{thebibliography}{10}
\bibitem{thy} M. B\"uttiker, Y. Imry, and R. Landauer, Phys. Lett. A {\bf 96}, 365 (1985);
H.-F. Cheung, E.K. Riedel, and Y. Gefen, Phys. Rev. Lett. {\bf 62}, 587 (1989); V. Ambegaokar and U. Eckern, Phys. Rev. Lett. {\bf 65}, 381 (1990); A. Schmid, Phys. Rev. Lett. {\bf 66}, 80 (1991); F. von Oppen and E.K. Riedel, Phys. Rev. Lett. {\bf 66}, 84 (1991); B.L. Altshuler, Y. Gefen, and Y. Imry, Phys. Rev. Lett. {\bf 66}, 88 (1991).
\bibitem{exp} L.P. Levy, G. Dolan, J. Dunsmuir, and H. Bouchiat, Phys. Rev. Lett. {\bf 64}, 2074 (1990); V. Chandrasekhar, R.A. Webb, M.J. Brady, M.B. Ketchen, W.J. Gallagher, and A. Kleinsasser, Phys. Rev. Lett. {\bf 67}, 3578 (1991); E.M.Q. Jariwala, P. Mohanty, M.B. Ketchen, and R.A. Webb  Phys. Rev. Lett. {\bf 86}, 1594 (2001); A.C. Bleszynski-Jayich, W.E. Shanks, B. Peaudecerf, E. Ginossar, F. von Oppen,
     L. Glazman, and J.G.E. Harris, Science {\bf 326}, 272 (2009).
\bibitem{SZ10} A.G. Semenov and A.D. Zaikin,  J. Phys.: Condens. Matter {\bf 22}, 485302 (2010); J. Phys.: Conf. Ser. {\bf 248}, 012034 (2010).
\bibitem{Buttiker} P. Cedraschi, V.V. Ponomarenko, and M. B\"uttiker,
Phys. Rev. Lett. {\bf 84}, 346 (2000); Ann. Phys. {\bf 289}, 1 (2001).
\bibitem{GZ98} D.S. Golubev and A.D. Zaikin, Physica B {\bf 255}, 164 (1998).
\bibitem{Paco} F. Guinea, Phys. Rev. B {\bf 65}, 205317 (2002).
\bibitem{GHZ} D.S. Golubev, C.P. Herrero, and A.D. Zaikin, Europhys. Lett.
{\bf 63}, 426 (2003).
\bibitem{AGZ} K.Yu. Arutyunov, D.S. Golubev, and A.D. Zaikin,
Phys. Rep. {\bf 464}, 1 (2008).
\bibitem{FH} R.P. Feynman and A.R. Hibbs, {\it Quantum Mechanics and Path
Integrals} (McGraw Hill, NY, 1965).
\bibitem{GZ1} D.S. Golubev and A.D. Zaikin, Phys. Rev. Lett. {\bf 81}, 1074 (1998); Phys. Rev. B {\bf 59}, 9195 (1999); Phys. Rev. B {\bf 62}, 14061 (2000); J. Low. Temp. Phys. {\bf 132}, 11 (2003).
\bibitem{GZS} D.S. Golubev, A.D. Zaikin, and G. Sch\"on, J. Low. Temp. Phys. {\bf 126}, 1355 (2002).
\bibitem{GZ2} D.S. Golubev and A.D. Zaikin, New J. Phys. {\bf 10}, 063027 (2008); Physica E {\bf 40}, 32 (2007).
\bibitem{pa91} S.V. Panyukov and A.D. Zaikin, Phys. Rev. Lett. {\bf 67}, 3168 (1991); J. Low Temp. Phys. {\bf 73}, 1 (1988).
\bibitem{HSZ} C.P. Herrero, G. Sch\"on, and A.D. Zaikin, Phys. Rev. B
{\bf 59}, 5728 (1999) and further references therein.
\bibitem{HlD} B. Horovitz and P. Le Doussal, Phys. Rev. B {\bf 74}, 073104 (2006); {\bf 82}, 155127 (2010).
\bibitem{CH} D. Cohen and B. Horovitz, J. Phys. A: Math. Theor. {\bf 40}, 12281 (2007);
  Europhys. Lett. {\bf 81}, 30001 (2008).
\bibitem{KH} V. Kagalovsky and B. Horovitz, Phys. Rev. B {\bf 78}, 125322 (2008).
\bibitem{SZ09} A.G. Semenov and A.D. Zaikin, Phys. Rev. B {\bf 80}, 155312 (2009).
\bibitem{GZ94} D.S. Golubev and A.D. Zaikin, Phys. Rev. B {\bf 50}, 8736 (1994).
\bibitem{SS} H. Schoeller and G. Sch\"on, Phys. Rev. B {\bf 50}, 18436 (1994).
\bibitem{Schmid} A. Schmid, J. Low Temp. Phys. {\bf 49}, 609 (1982).
\bibitem{AES} U. Eckern, G. Sch\"on, and V. Ambegaokar, Phys. Rev. B {\bf 30}, 6419 (1984).
\bibitem{GZ92} D.S. Golubev and A.D. Zaikin, Phys. Rev. B {\bf 46}, 10903 (1992); Phys. Rev. Lett. {\bf 86}, 4887 (2001).
\bibitem{SZ11} A.G. Semenov and A.D. Zaikin, Phys. Rev. B {\bf 84}, 045416 (2011).
\bibitem{GZ96} D.S. Golubev and A.D. Zaikin, JETP Lett. {\bf 63}, 1007 (1996);
D. Chouvaev, L.S. Kuzmin, D.S. Golubev, and A.D. Zaikin, Phys. Rev. B {\bf 59}, 10599 (1999).
\bibitem{ZGOZ} A.D. Zaikin, D.S. Golubev, A. van Otterlo, and G.T. Zimanyi, Phys. Rev. Lett. {\bf 78}, 1552 (1997).
\bibitem{GZ01} D.S. Golubev and A.D. Zaikin, Phys. Rev. B {\bf 64}, 014504 (2001).
\bibitem{anne} A. van Otterlo, D.S. Golubev, A.D. Zaikin, and G. Blatter, Eur. Phys. J. B {\bf 10}, 131 (1999).
\end{thebibliography}
\end{document}